\documentclass[
reprint,
superscriptaddress,
 amsmath,
 amssymb,
 aps,
prl,
]{revtex4-1}

\usepackage{graphicx}
\usepackage{dcolumn}
\usepackage{bm}

\usepackage{color}
\usepackage{colordvi}  
\definecolor{Red}{rgb}{1,0,0}

\makeatletter
\def\authornote{\xdef\@thefnmark{$\dagger$}\@footnotetext}
\makeatother

\begin{document}

\title{Coherent generation of nonclassical light on chip via detuned photon blockade}

\author{Kai M\"uller$^\dagger$}
\email{kaim@stanford.edu}
\affiliation{E. L. Ginzton Laboratory, Stanford University, Stanford, California 94305, USA}
\author{Armand Rundquist$^\dagger$}
\affiliation{E. L. Ginzton Laboratory, Stanford University, Stanford, California 94305, USA}
\authornote{These authors contributed equally.}
\author{Kevin A. Fischer$^\dagger$}
\affiliation{E. L. Ginzton Laboratory, Stanford University, Stanford, California 94305, USA}
\author{Tomas Sarmiento}
\affiliation{E. L. Ginzton Laboratory, Stanford University, Stanford, California 94305, USA}
\author{Konstantinos G. Lagoudakis}
\affiliation{E. L. Ginzton Laboratory, Stanford University, Stanford, California 94305, USA}
\author{Yousif A. Kelaita }
\affiliation{E. L. Ginzton Laboratory, Stanford University, Stanford, California 94305, USA}
\author{Carlos S\' anchez Mu\~noz}
\affiliation{Departamento de F\'isica Te\'orica de la Materia Condensada and Condensed Matter Physics Center (IFIMAC), Universidad Aut\'onoma de Madrid, E-28049, Spain}
\author{Elena del Valle}
\affiliation{Departamento de F\'isica Te\'orica de la Materia Condensada and Condensed Matter Physics Center (IFIMAC), Universidad Aut\'onoma de Madrid, E-28049, Spain}
\author{Fabrice P. Laussy}
\affiliation{Departamento de F\'isica Te\'orica de la Materia Condensada and Condensed Matter Physics Center (IFIMAC), Universidad Aut\'onoma de Madrid, E-28049, Spain}
\affiliation{Russian Quantum Center, Novaya 100, 143025 Skolkovo, Moscow Region, Russia}
\author{Jelena Vu\v{c}kovi\'c}
\affiliation{E. L. Ginzton Laboratory, Stanford University, Stanford, California 94305, USA}

\date{\today}

\begin{abstract}
The on-chip generation of non-classical states of light is a key-requirement for future optical quantum hardware. In solid-state cavity quantum electrodynamics, such non-classical light can be generated from self-assembled quantum dots strongly coupled to photonic crystal cavities. Their anharmonic strong light-matter interaction results in large optical nonlinearities at the single photon level, where the admission of a single photon into the cavity may enhance (photon-tunnelling) or diminish (photon-blockade) the probability for a second photon to enter the cavity. Here, we demonstrate that detuning the cavity and QD resonances enables the generation of high-purity non-classical light from strongly coupled systems. For specific detunings we show that not only the purity but also the efficiency of single-photon generation increases significantly, making high-quality single-photon generation by photon-blockade possible with current state-of-the-art samples.
\end{abstract}

\pacs{Valid PACS appear here}
\maketitle

Due to their strong interaction with light and ease of integration into optoelectronic devices, self-assembled quantum dots (QDs) are promising candidates for quantum light sources \cite{2014_Chang}. High-fidelity single-photon generation from QDs for off-chip applications has been demonstrated under both non-resonant \cite{2001_Santori} and resonant \cite{2012_Matthiesen, 2013_He, 2009_Ates} excitation. Some of these experiments have employed micro-pillar cavities \cite{2013_Gazzano}, etched \cite{2010_Claudon} or epitaxially grown photonic nanowires \cite{2012_Reimer} for enhanced light off-chip extraction efficiency. On the other hand, photonic crystal cavities provide a promising on-chip route toward optoelectronic integration of QDs due to the established set of associated integrated waveguide and detector structures \cite{2013_Reithmaier,2011_Sprengers}. Such structures will be able to exploit strong light-matter coupling with QDs for the generation of a variety of on-chip non-classical light states by various quantum-electrodynamical (QED) methods, and recent exotic proposals have even explored the possibility of releasing energy exclusively in bundles of $n$-photons \cite{2014_Laussy}. The phenomena of photon-tunnelling and photon-blockade in strongly coupled systems have been experimentally demonstrated both for the case of the QD on resonance \cite{2008_Faraon_Blockade, 2010_Faraon_Blockade, 2012_Majumdar_Ladder} and near resonance \cite{2011_Reinhard_Correlated} with the cavity (and likewise, only for resonant atom-cavity system \cite{Birnbaum_2005}). However, in the case of large detuning these effects have only been investigated theoretically \cite{2012_Laussy_Climbing}.

In this letter, we demonstrate the feasibility of performing photon-blockade at significant detuning, and indeed the importance of doing so for high-purity and high-efficiency operation. We show that by detuning the QD and cavity resonances while operating in the photon-blockade regime, the second-order autocorrelation function ($g^{(2)}(0)$) of the light transmitted through the cavity decreases from $g^{(2)}(0)=0.9 \pm 0.05$ to $g^{(2)}(0)=0.29 \pm 0.04$. Simulations of the second- and third-order autocorrelation functions for our system are in excellent agreement with the measurements, and they reveal that not only does the quality of the single photon stream increase, but that the absolute probability of obtaining a single photon increases by a factor of $\sim 2$. Furthermore, we show that the values we obtain for $g^{(2)}(0)$ are only limited by the system parameters (QD-cavity field coupling strength $g$ and cavity field decay rate $\kappa$), and that high-quality single-photon emission is within reach for current state-of-the-art samples for specific cavity and QD detunings.

\begin{figure}[!t]
  \includegraphics[width=\columnwidth]{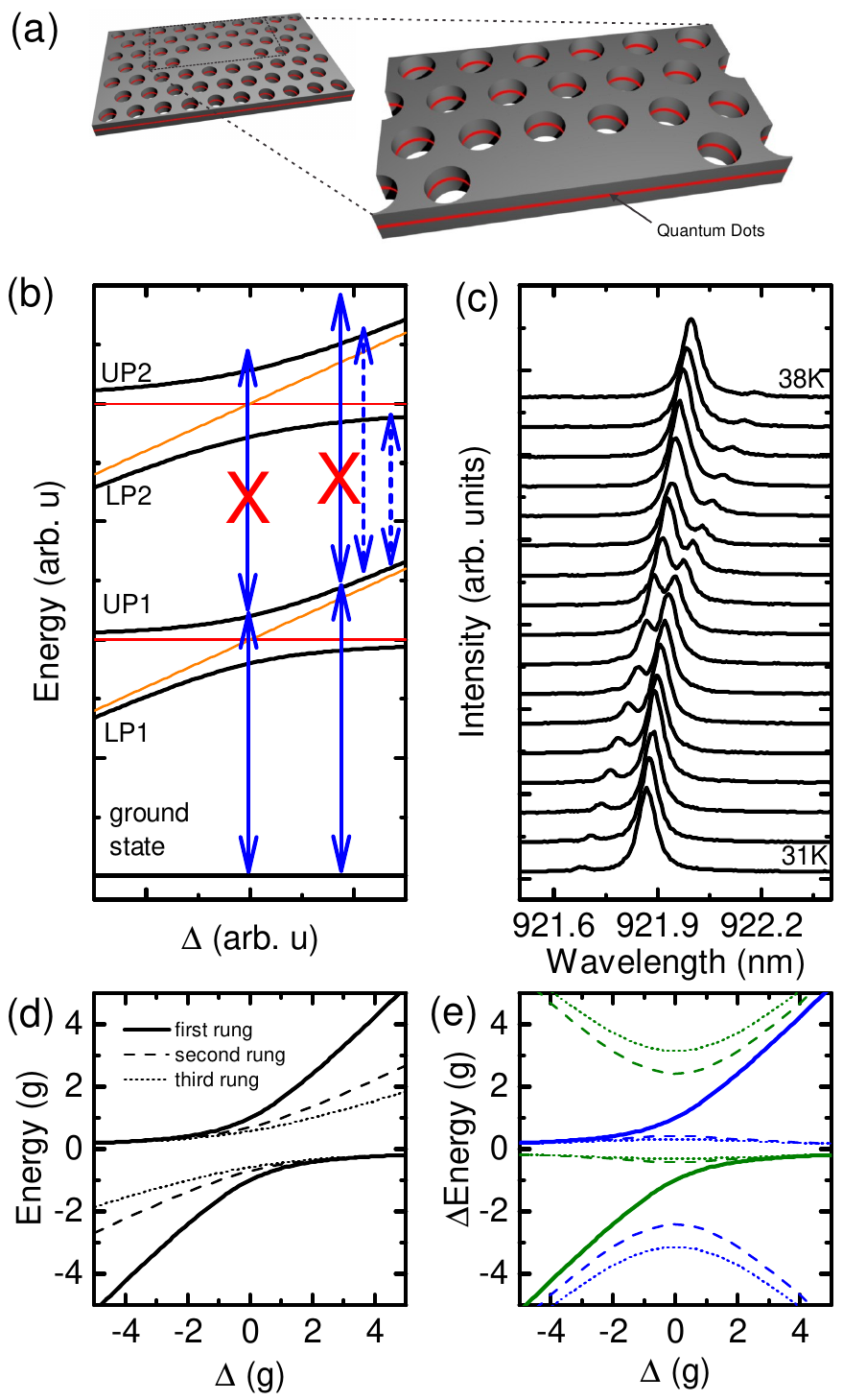}
  \caption{(a) Schematic illustration of self-assembled QDs embedded in a photonic crystal cavity. (b) Jaynes-Cummings ladder obtained from equation \ref{equation:1} (c) Cross-polarised reflectivity spectrum of the coupled QD-cavity system obtained for tuning the QD through the cavity resonance. (d) Energies for exciting the $n$-th rung of the Jaynes-Cummings ladder in an $n$-photon process. (e) Transient energies for climbing the Jaynes-Cummings-ladder rung by rung. Transitions from upper and lower polaritons are colour coded in blue and green, respectively. In panels (d) and (e) the energy of the bare cavity was subtracted from all transitions for better comparison. $\Delta$ is the QD-cavity detuning and $g$ the coupling strength.}
  \label{figure:1}
\end{figure}

The sample investigated is schematically illustrated in figure \ref{figure:1}a and consists of a layer of low density InAs QDs grown by molecular beam epitaxy and embedded in a photonic crystal L3 cavity \cite{2003_Akahane}. The energy structure of a QD strongly coupled to a cavity is well described by the Jaynes-Cummings (JC) Hamiltonian
\begin{equation}
H = \omega_a a^{\dagger} a + (\omega_ a + \Delta)\sigma^{\dagger}\sigma + g(a^{\dagger}\sigma + a\sigma^{\dagger})
\label{equation:1}
\end{equation}
where $\omega_a$ denotes the frequency of the cavity, $a$ the annihilation operator associated with the cavity mode, $\sigma$ the lowering operator of the quantum emitter, $\Delta$ the detuning between quantum emitter and cavity, and $g$ the emitter-cavity field coupling strength. The resulting eigenenergies, the Jaynes-Cummings-ladder dressed states, are illustrated in figure \ref{figure:1}b. For $n$ photons in the cavity the energy is $n \omega_a$ (red lines), and the energy of the quantum emitter (orange) varies with a detuning parameter. Due to the coupling, the resulting energy eigenstates are the anticrossing polariton branches. At resonance, the splitting is given by $2g\sqrt{n}$ (with $n$ being the index of the rung). While this letter explicitly discusses the case of a QD in a photonic crystal cavity, the same physics holds for a large number of systems such as those formed by atoms \cite{1996_Brune, Schuster_2008} or superconducting circuits \cite{2008_Fink}.

For QDs, the anticrossing that results from the coupling to a cavity can be efficiently studied in optical spectroscopy experiments, where the QD and cavity detuning is controlled by the lattice temperature \cite{2004_Reithmaier, 2007_Hennessy}. The result of such a measurement is presented in figure \ref{figure:1}c, which shows the transmission through the cavity measured in a cross-polarised reflectivity configuration \cite{2007_Englund} in the temperature range $T=31-38 \, \text{K}$. A clear anticrossing provides evidence of strong coherent coupling between QD and cavity. A fit (not shown here) reveals a coupling strength of $g/2\pi = 10.9 \, \text{GHz}$ and a cavity field decay rate $\kappa /2\pi = 10.0 \, \text{GHz}$.

Due to the unequal energy spacing (anharmonicity) of the Jaynes-Cummings ladder, transmission of a laser through the cavity affects the beam's photon statistics and introduces strong – photon correlations \cite{2008_Faraon_Blockade, 2011_Reinhard_Correlated}. This is schematically illustrated by the solid blue arrows in figure \ref{figure:1}b; if the laser is tuned into resonance with one of the polariton branches of the first rung,  it cannot excite the system to the second rung due to the ladder anharmonicity. Therefore, in this regime the transmitted beam consists of a series of single photons and hence is called the photon-blockade regime. However, the fidelity of this process is inherently limited by the transition linewidth, given by the cavity field $\kappa$ and quantum emitter $\gamma$ decay rates. In particular, due to final state broadening and the shorter lifetime of excited states, transitions to higher rungs have larger linewidths, further reducing the probability of generating single photons.

Importantly, operating the system at a significant QD-cavity detuning can lead to higher-purity single-photon emission. We consider two cases to support this conclusion: the excitation of a higher rung in a multi-photon process and subsequent excitation. Therefore, we plot the energies for an $n$-photon excitation of the $n$-th rung in figure \ref{figure:1}d and the transient energies from one rung to the next in figure \ref{figure:1}e. Clearly, at zero detuning the energies for exciting the first and higher rungs are close together (figure \ref{figure:1}d), and their separation strongly increases for the upper (lower) polariton branch for positive (negative) detunings of the quantum emitter. For a laser in resonance with the first rung the probability of $n$-photon excitation of higher rungs decreases with increased detuning. Similar scenarios can be found for subsequent climbs up the ladder, as presented in figure \ref{figure:1}e, which shows the transition energies from the ground state to the first rung, the first to the second rung and the second to the third rung as solid, dashed and dotted lines, respectively. Transitions from an upper (lower) polariton branch to higher rungs are colour coded in blue (green). Near resonance the first and second transitions are close in energy but their separation strongly increases with the detuning of the quantum emitter (c.f. blue arrows in figure \ref{figure:1}b). The close proximity of the first rung to the outer higher order transitions for large detunings does not reduce the single-photon emission character, since these transitions occur from the other polariton branch as can be seen from the different colours. Therefore, a detuning between quantum emitter and cavity is also expected to improve the purity of single-photon generation under photon-blockade for subsequent rung excitation. Furthermore, detuning also affects the linewidths of the states in such a way that the linewidth of a polariton branch that evolves towards the bare QD (bare cavity) transition decreases (increases). This further reduces the overlap of transitions involving different rungs of the JC ladder and increases the fidelity of photon-blockade (see supplemental material).

\begin{figure}[!t]
  \includegraphics[width=\columnwidth]{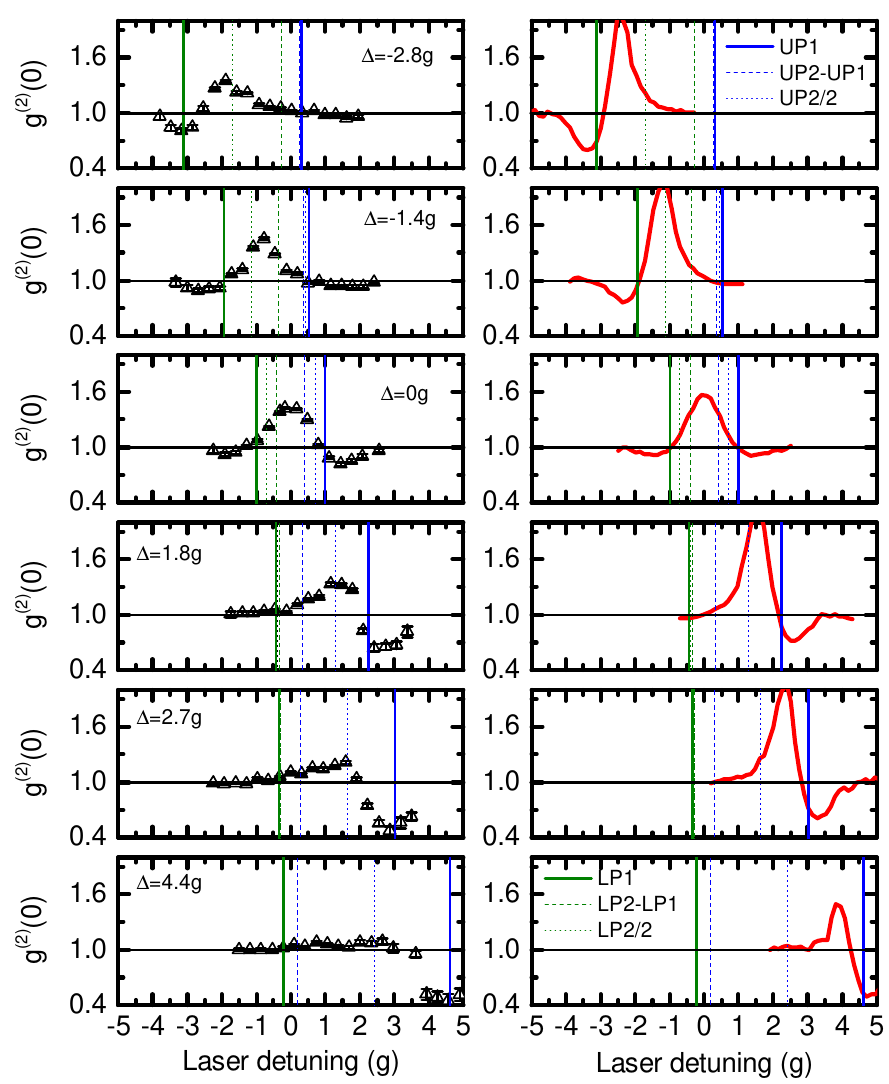}
  \caption{$g^{(2)}(0)$ as function of the laser detuning for a set of different QD-cavity detunings: (left) experiment and (right) simulation. With increased detuning the depth of the anti-bunching is more pronounced. Vertical lines represent the relevant transition energies of the JC ladder as described in figure \ref{figure:1}.}
  \label{figure:g2}
\end{figure}

To quantify the quantum character of light the second-order autocorrelation function \cite{1963_Glauber,2000_Loudon}
\begin{equation}
g^{(2)}(0) = \frac{\left\langle m(m-1) \right\rangle}{\left\langle m \right\rangle^2}
\label{equation:2}
\end{equation}
is a commonly used quantity, where $m$ signifies a number of detections in the photocount distribution. It results in a $g^{(2)}(0)$ of 1 for a coherent source and 0 for a perfect stream of single photons. To test our expectation that the purity of single-photon generation under photon-blockade can be improved by detuning the QD and cavity resonances, we measured $g^{(2)}(0)$ from the output correlations of a laser beam transmitted through the cavity. The result of these experiments is presented in the left part of figure \ref{figure:g2} that shows $g^{(2)}(0)$ as a function of the laser detuning for six different QD and cavity detunings. The data were recorded under pulsed excitation with $t_p = 30 \, \text{ps}$ long pulses. This pulse duration was chosen as a compromise between frequency resolution and avoiding re-excitation of the system. In the case of  $\Delta \approx 0$, the form of $g^{(2)}(0)$ is nearly symmetric with photon tunnelling generating a maximum of $g^{(2)}(0)=1.45 \pm 0.05$ in the centre, and photon-blockade generating a minimum dip of $g^{(2)}(0)=0.85 \pm 0.05$ ($g^{(2)}(0)=0.92 \pm 0.05$) at the laser detuning of $1.5g$ ($-1.5g$). When detuning the QD, the maximum of $g^{(2)}(0)$ shifts such that it stays between the polariton branches before it disappears for detunings greater than $\sim 4g$. The dip of $g^{(2)}(0)$ both moves with and shifts toward the polariton branch that is closer to the bare QD transition. Most strikingly, the depth of the dip increases and reaches a value as low as $g^{(2)}(0)=0.45 \pm 0.05$ for the detunings of $\Delta = 2.7g$ and $\Delta = 4.4g$. This value is lower than $0.5$, indicative of strong single-photon character, and lower than $g^{(2)}(0)$ measured in any prior photon-blockade experiments in the solid state. We note here that since the lifetime of the polariton branch closer to the bare QD transition increases with detuning (for details see supplemental material), excitation with $70 \, \text{ps}$ long pulses was possible without re-exciting the system at detunings of $\Delta = 3-5 \, g$, further reducing antibunching to $g^{(2)}(0)=0.29 \pm 0.04$ (see supplemental material). Small asymmetries in the experimental measurements result from the wavelength dependence of the cross-polarized laser suppression, asymmetries in the spectral shape of the laser pulse, drift of the QD-cavity detuning, and temperature tuning between curves.

To support our findings, we performed quantum optical simulations using a quantum trajectory method (see supplemental material). The results of these simulations are presented on the right side of figure \ref{figure:g2}. Overall, the simulations are in excellent qualitative agreement with the measurements and also quantitatively resemble the values measured in the photon-blockade regime. Only small differences exist: the measured maximum values of $g^{(2)}(0)$ are slightly lower than the simulated ones. This can be explained by blinking of the quantum emitter \cite{2011_Reinhard_Correlated}, which was not included in the simulations.

\begin{figure}[!t]
  \includegraphics[width=\columnwidth]{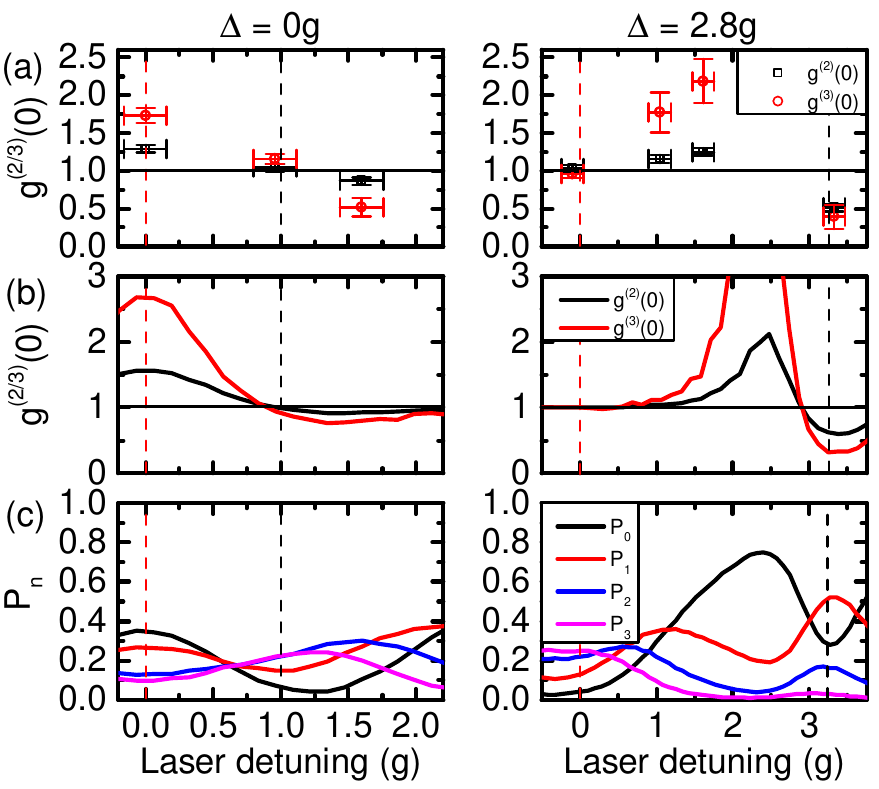}
  \caption{(a) Measured second- and third-order autocorrelation functions as a function of the laser detuning for a QD detuning of $\Delta=0$ (left) and $\Delta=2.8g$ (right). (b) Simulation for the experimental conditions presented in (a). (c) Simulated probabilities for having $n=0-3$ photons in the output of a pulse transmitted through the cavity. Clearly, for a detuning of $\Delta=2.8g$ (right) $P_1$ (red) exhibits a pronounced peak. Vertical dashed lines represent the bare cavity (red) and UP1 (black) frequencies.
	}
  \label{figure:g3}
\end{figure}

To further investigate the single-photon character of the light transmitted through the cavity we performed measurements of the third-order autocorrelation function $g^{(3)}(0) = \frac{\left\langle m(m-1)(m-2) \right\rangle}{\left\langle m \right\rangle^3}$, as higher-order autocorrelations are necessary to characterise the multi-photon nature of non-classical light \cite{2014_Rundquist}. The result of these measurements are presented in figure \ref{figure:g3}a, which shows $g^{(2)}(0)$ and $g^{(3)}(0)$ as a function of the laser detuning for the case of QD-cavity detuning of $\Delta = 0$ (left) and $\Delta = 2.8g$ (right). Clearly, $g^{(3)}(0)$ shows the same qualitative shape as $g^{(2)}(0)$ but with stronger non-classical values. Simulations of these autocorrelations are presented in figure \ref{figure:g3}b and show good agreement with the measurements. In particular, for the photon-blockade regime the values of $g^{(3)}(0)$ are lower than those of $g^{(2)}(0)$, indicating that $g^{(2)}(0)$ is mainly limited by two-photon events and not higher photon events. 

Since the agreement with the measured autocorrelation functions is very good, we can rely on the simulations to explicitly access quantities only within reach of the theory, such as the probabilities $P_n$ of transmitting $n$ photons per excitation pulse through the cavity. These probabilities are presented in figure \ref{figure:g3}c for $n=0-3$ under the same conditions as the data presented in figure \ref{figure:g3}a and \ref{figure:g3}b. Interestingly, we find that in the case of zero QD and cavity detuning (figure \ref{figure:g3}c left), we see significant contributions of one, two and three-photon events for all laser detunings. In fact, the probability for two-photon events (blue) actually dominates over the probability for single photons (red) in the case of the best photon-blockade. In strong contrast, for a QD-cavity detuning of $\Delta = 2.8g$ (figure \ref{figure:g3}c right) and operation in the photon-blockade regime, single-photon events (red) strongly dominate over two-photon events (blue) and the probability for three-photon events (purple) becomes negligible. Most strikingly, in the detuned case, not only does the quality of the single-photon stream increase, but the absolute probability of finding a single photon in the transmitted laser pulse increases by a factor of $\sim 2$. In addition to the agreement between measured and simulated values for both $g^{(2)}(0)$ and $g^{(3)}(0)$, the experimental count rates support this finding. Since the overall count rate is proportional to $\sum n P_n$, it does not directly correspond to the overall single-photon efficiency. However, we can still calculate the ratio of the count rates at different detuning conditions in order to compare simulation and experiment. For the points of best photon blockade in the resonant and detuned case, our simulated count rates result in a ratio of 2.27 : 1.05 = 2.16, which is in very good agreement with the measured ratio of $4\times 10^4$ : $1.8\times 10^4$ = 2.22.

This counter-intuitive finding that the efficiency of single-photon generation increases when detuning cavity and QD can be understood in the following way: Photon-blockade is obtained if the first rung of the JC ladder is excited while the overlap of the laser with higher rungs is suppressed. When on-resonance this suppression is inherently limited by the linewidth of the transitions, that scales with $n$ as the decay rate of a rung is proportional of the number of photons. Meanwhile, the detuning of subsequent rungs scales with $\sqrt{n}$ (see supplemental material). Therefore, for any system parameters $\kappa$ and $g$ that can be achieved with the emitter and cavity in resonance, there will always be an overlap between the transition to the first rung and to higher climbs up the ladder. As a result, the strongest photon-blockade with the emitter and cavity in resonance is not observed for the laser exactly on resonance with the first rung of the JC-ladder ($\sim \pm g$), but rather with the laser off-resonant and detuned to $\sim \pm 1.5 \, g$ (c.f. figure \ref{figure:g3} - left). In contrast, if the separation between different JC rungs is enhanced by detuning the emitter and cavity, the strongest photon-blockade is obtained with the laser resonant with the polariton branch, making photon-blockade more efficient than in the resonant case. Therefore, not only the purity but also the efficiency of single-photon generation improves given the correct detuning between cavity and emitter. With increasing detuning between the QD and cavity, the oscillator strength of the more QD-like polariton branch decreases as the oscillator strength of the QD is much weaker than the one of the cavity. Therefore, for too large detunings the efficiency decreases, resulting in an \emph{optimum detuning} for single-photon generation of a few $g$ (see supplemental material).

This approach to photon-blockade has strong potential for single-photon generation under already achievable system parameters. Improvements in the spatial alignment of the QD and cavity field have enabled the coupling strength to reach values up to $g/2\pi = 40 \, \text{GHz}$ \cite{2013_Takamiya}. Recent nanofabrication improvements have allowed for experimental GaAs photonic crystal cavity loss rates as low as $\kappa /2\pi = 4.0 \, \text{GHz}$\cite{2011_Ota}. When using these parameters in our simulations we obtain $g^{(2)}(0) = 0.1$ in the photon-blockade regime, and an absolute probability of over $90 \%$ for single-photon emission, demonstrating that high-quality single-photon streams generated by photon-blockade are within reach.

In summary, we have demonstrated that QD-cavity detuning is a key ingredient for high-purity generation of non-classical light from strongly coupled systems. We have shown that detuning strongly reduces the spectral overlap with higher rungs of the Jaynes-Cummings-ladder and hence greatly improves the generation of single photons by photon-blockade. We have presented quantum-optical simulations that are in excellent agreement with our measurements and show that high-quality single-photon generation under photon-blockade is possible with current state-of-the-art samples. The generation of single photons by photon-blockade might have advantages over other techniques. First, the use of high quality photonic crystal cavities promises a method of on-chip routing of the photons by coupling them to photonic crystal waveguides (with high efficiency) \cite{Faraon_2008}. Second, the cavity emission rate is at least one order of magnitude faster than the bare QD emission rate, resulting in a comparable increase in the maximum single-photon generation rate while maintaining potential advantages from resonant excitation. Furthermore, the successful experimental demonstration of photon-blockade in the detuned light-matter configuration demonstrates the feasibility of operating cavity QED in such an extreme regime and paves the way for a wealth of other quantum light sources, including those generating $n$-photon states \cite{2014_Laussy}.

We gratefully acknowledge financial support from the Air Force Office of Scientific Research, MURI Center for Multifunctional light-matter interfaces based on atoms and solids (Grant No. FA9550-12-1-0025) and support from the Army Research Office (grant number W911NF1310309). KM acknowledges financial support from the Alexander von Humboldt Foundation. KGL acknowledges financial support from the Swiss National Science Foundation. KAF acknowledges support through Lu Stanford Graduate Fellowship. YAK acknowledges support through Stanford Graduate Fellowship and National Defense Science and Engineering Graduate Fellowship. FPL acknowledges support from the ERC grant POLAFLOW.



\providecommand{\noopsort}[1]{}\providecommand{\singleletter}[1]{#1}%

\cleardoublepage

\onecolumngrid
\section*{Supplemental Material}
\twocolumngrid

\section*{Methods}
\textbf{Sample fabrication}\\
The MBE grown structure consists of a $\sim 900 \, \text{nm}$ thick Al$_{0.8}$Ga$_{0.2}$As sacrificial layer followed by a $145 \, \text{nm}$ thick GaAs layer that contains a single layer of InAs QDs. Our growth conditions result in a typical QD density of $60-80 \, \mu \text{m}^{-2}$. The photonic crystals were fabricated using $100 \, \text{keV}$ e-beam lithography with  ZEP resist, followed by reactive ion etching and HF removal of the sacrificial layer. The photonic crystal lattice constant was $a = 246 \, \text{nm}$ and the hole radius $r \approx 60 \, \text{nm}$. The cavity fabricated is a linear three-hole defect (L3) cavity. To improve the cavity quality factor, holes adjacent to the cavity were shifted \cite{2005_Akahane, 2014_Minkov}.
\\

\textbf{Optical spectroscopy}\\
All optical measurements were performed with a liquid helium flow cryostat at temperatures in the range $10-40 \, \text{K}$. For excitation and detection a microscope objective with a numeric aperture of $NA = 0.75$ was used. Cross-polarised measurements were performed using a polarising beam splitter. To further enhance the extinction ratio, additional thin film linear polarisers were placed in the excitation/detection pathways and a single mode fibre was used to spatially filter the detection signal. Furthermore, two waveplates were placed between the beamsplitter and microscope objective: a half-wave plate to rotate the polarisation relative to the cavity and a quarter-wave plate to correct for birefringence of the optics and sample itself.
\\

\textbf{Autocorrelation measurements}\\
Second-order autocorrelation measurements were performed using a Hanbury Brown and Twiss (HBT) setup consisting of one fibre beamsplitter and two single photon avalanche diodes. The  detected photons were correlated with a PicoHarp300 time counting module. Third order autocorrelation measurements were performed using a generalised HBT setup consisting of three fibre beamsplitters that result in a balanced splitting of the signal into four channels and four single photon avalanche diodes. In the generalised setup, the detected photons were correlated using a sensL four channel multi-stop time-tagging counting module. The excitation was performed using a frequency filtered Ti:Sapphire laser with a repetition rate of 80MHz. For measurements of $g^{(2)}(0)$ the count rate per detector ranged from $1 \cdot 10^4$ to $1 \cdot 10^5$ counts per second, depending on the laser and QD-cavity detuning. Therefore, the integration time per datapoint ranged from $100s$ to $1800s$. For the measurements of $g^{(3)}(0)$ the integration time was $5-10$ hours per datapoint.\\

\section*{Relation between $g^{(2)}(0)$ and correlation measurements}
The expression $g^{(2)}(0) = \frac{\langle a^{\dagger} a^{\dagger} a a \rangle}{\langle a^{\dagger} a \rangle^2}$ is commonly used to describe the second-order field autocorrelation of a given light beam. Following Loudon \cite{2000_Loudon}, we first discuss the validity of this expression as used in the context of our pulsed excitation scheme and mention that it is somewhat incomplete. A semiclassical treatment of photodetection defines the measured degree of second-order coherence at zero time delay as $g^{(2)}(0) \equiv \frac{\langle m(m-1) \rangle}{\langle m \rangle^2}$, which is the normalized second order factorial moment of the time integrated photocount distribution $P_m (T)$. In analogy to the classical integrated mean intensity, the quantum mechanical operator
\begin{equation}
\hat{M} (T) =\int_0^T dt \, b^{\dagger}(t) b(t)
\end{equation}
represents the total photon number arriving at an ideal detector over the time interval $t\in[0,T]$, such that $\langle m \rangle = \langle \hat{M}(T) \rangle$〉. Here, $b(t)$ is the instantaneous field mode operator describing the field flux - it is important to note its distinction with the cavity field mode operator $a(t)$: $b(t)$ describes a field flux while $a(t)$ describes a field. Comparing our semiclassical definition of $g^{(2)}(0)$ to the quantum mechanical operator $\hat{M}(T)$, we have
\begin{equation}
g^{(2)}(0) = \frac{\langle \hat{M}(T)(\hat{M}(T)-1) \rangle}{\langle \hat{M}(T)\rangle^2} = \frac{\langle: \hat{M}(T)^2:\rangle}{\langle \hat{M}(T) \rangle^2}
\end{equation}
 where $\langle: \hat{M}^2(T) :\rangle$ denotes the quantum mechanical expectation value of the normally ordered second moment of the photon number operator. Writing out this expression explicitly
\begin{equation}
g^{(2)}(0) = \frac{\int_0^T \int_0^T dt dt' \, \langle b^{\dagger}(t) b^{\dagger}(t') b(t') b(t) \rangle}{(\int_0^T dt \, \langle b^{\dagger}(t) b(t) \rangle)^2}
\label{eq:g2x}
\end{equation}
and substituting the un-normalized correlation moments for the expectation values, we arrive at
\begin{equation}
g^{(2)}(0)= \frac{\int_0^T \int_0^T dt dt' \, G_{bb}^2 (t,t')}{\langle m \rangle^2} 
\end{equation}
Therefore, the measured $g^{(2)}(0)$ actually represents a quantity other than that which $\frac{\langle a^{\dagger} a^{\dagger} a a \rangle}{\langle a^{\dagger} a \rangle^2}$  would suggest. Instead of representing the correlation between two photon number (Fock) states - states that incidentally do not exist in free space - $g^{(2)}(0)$ represents the sum of all field flux correlations $G_{bb}^2 (t,t')$ over the detection time. In other words, $\frac{\langle a^{\dagger} a^{\dagger} a a \rangle}{\langle a^{\dagger} a \rangle^2}$ operates only on states which are well defined in a cavity whereas the correlation measurements are performed on fluxes. However, input-output theory provides a direct connection between the internal and external mode operators such that measuring zero delay flux correlations is equivalent to calculating $\frac{\int_0^T \int_0^T dt dt' \langle a^{\dagger}(t) a^{\dagger}(t') a(t') a(t) \rangle}{\left( \int_0^T dt \langle a^{\dagger}(t) a(t) \rangle \right)^2}$ inside the cavity.\\

\section*{Details on the simulations of $g^{(2)}(0)$ and $P_n$}
Assuming the rotating wave approximation (RWA) and a rotating reference frame relative to the laser field (frequency $\omega_l$), the Jaynes-Cummings Hamiltonian describing interaction of our strongly coupled system with a laser pulse takes the form
\begin{multline}
H= \Delta_{QD} \sigma_+ \sigma_- + \Delta_a a^{\dagger} a \\ + g(a^{\dagger}\sigma_- + a \sigma_+) + \mathcal{E}(t)(a+a^{\dagger})
\end{multline}
where $\sigma_{\pm}$ are the dipole operators, $a$ is the cavity mode operator, and $\mathcal{E}(t)$ the slowly varying field-cavity coupling amplitude. The self-energies and their shifts in the rotating frame are defined by $\Delta_{QD} = \omega_{QD} - \omega_l$ and $\Delta_a = \omega_a - \omega_l$, where $\omega_{QD}$ and $\omega_a$ are the atomic and cavity energies, respectively. We take the laser pulse as a Gaussian, well described by $\mathcal{E}(t) = \mathcal{E}_0 \exp{(-\frac{t^2}{2\tau_p^2})}$, where $\tau_p = 30 \, \text{ps}$ is the pulse parameter. In simulations, we find that $\mathcal{E}_0$ results in correlations that closely match the observed experimental data. Additionally, we assume that the cavity decay, $\kappa$, dominates the loss dynamics because the quantum dot lifetime is almost two orders of magnitude larger than the cavity lifetime.

\begin{figure}[htb]
  \includegraphics[width=\columnwidth]{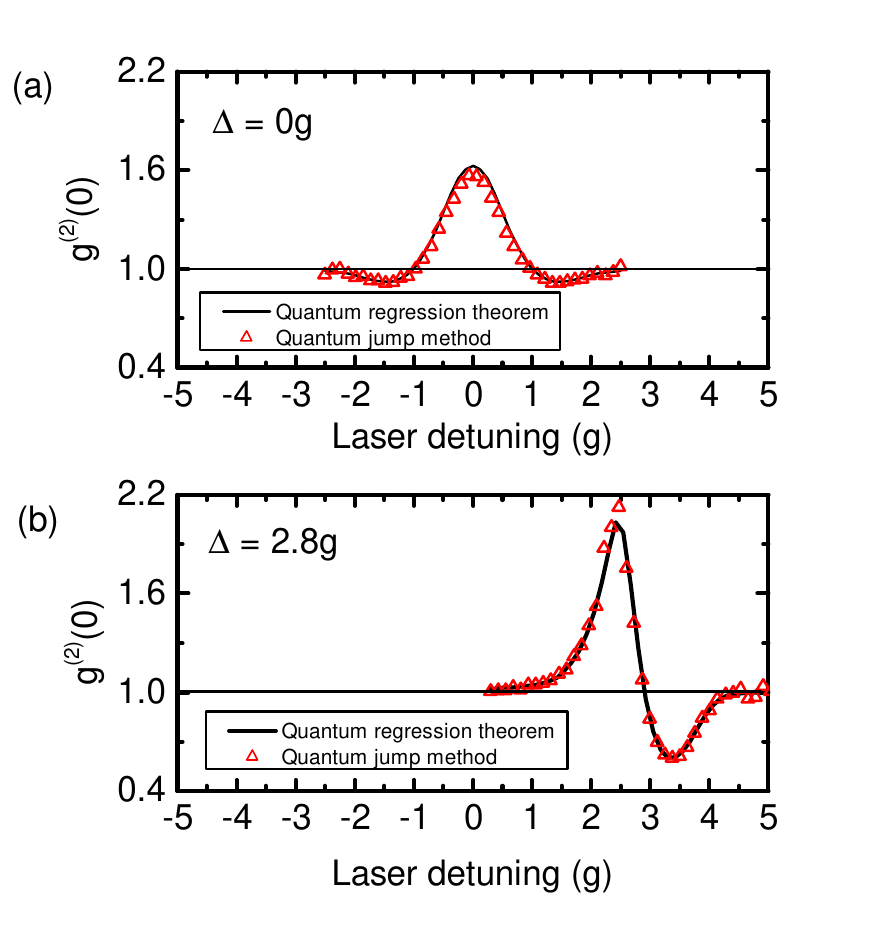}
  \caption{Simulated $g^{(2)}(0)$ obtained using the quantum regression theorem (solid lines) and quantum jump method (red data points) for a QD-cavity detuning of (a) $\Delta = 0$ and (b) $\Delta = 2.8g$	
	}
  \label{figure:S1}
\end{figure}

The simulations presented in this paper are performed using the Quantum Optics Toolbox in Python (QuTiP) \cite{2014_Johansson}. To simulate the second order correlation function two different approaches are used: the quantum regression theorem and the quantum jump method. Application of the quantum regression theorem to a problem of the form
\begin{equation}
\langle A(t) B(t+\tau)C(t)\rangle
\label{eq:qrt}
\end{equation}
yields the following result
\begin{equation}
\langle A(t) B(t+\tau)C(t)\rangle = \text{Tr}_{\text{sys}}\{B \Lambda (t, t+\tau) \}
\end{equation}
where $\Lambda(t, t+\tau)$ is governed by the following evolution equation 
\begin{equation}
\partial_r \Lambda(t,t+\tau) = \mathcal{L}(t+\tau) \Lambda(t,t+\tau) 
\end{equation}
subject to the initial condition
\begin{equation}
\Lambda(t,t) = C\rho(t)A
\end{equation}
After careful time-ordering \cite{Knoll1986}, equation \ref{eq:g2x} has the same form as equation \ref{eq:qrt} and thus the implementation of the quantum regression theorem for the calculation of $g^{(2)}(0)$ is straightforward.

In the quantum jump method, quantum trajectories that are solutions for our system's associated stochastic Schr\"odinger equation are calculated. Monitoring collapse events of the cavity field over many trajectories generates the expected photocount distribution that would be measured by an ideal, infinite bandwidth detector placed directly at the cavity output. From this distribution, we directly compute the normalised second order factorial moment in order to simulate our experimentally measured $g^{(2)}(0)=\frac{\left\langle m(m-1) \right\rangle}{\left\langle m \right\rangle^2}$.

While using the quantum regression theorem results in a faster simulation of $g^{(2)}(0)$, the quantum jump method additionally provides explicitly the probabilities for $n$ photons per pulse being transmitted through the cavity. To validate that the quantum jump method works well in our case and that we have simulated a sufficiently large number of quantum trajectories, we present in figure \ref{figure:S1} simulated values of $g^{(2)}(0)$ as a function of the laser detuning for two different QD-cavity detunings (c.f. figure 2 in the main text), as obtained from the quantum regression theorem and quantum jump method (black solid lines and red datapoints, respectively). Clearly, both methods produce excellent agreement.

Furthermore, we note that because the correlation functions describing photon statistics are robust to loss and our detectors have finite detection probability paired with low photon coincidence probability, the experimentally measured and normalized coincidences from the HBT setup should very closely match these simulated values.\\

For the simulations presented throughout this letter, we used the experimentally extracted cavity field decay rate $\kappa / 2 \pi= 10.0 \, \text{GHz}$ and coupling strength $g / 2 \pi= 10.9 \, \text{GHz}$. The QD decay rate $\gamma$ plays almost no role for the detunings investigated here as it is negligible due to a long intrinsic lifetime that is even further Purcell suppressed (as will be discussed in detail below). Therefore, the only parameter that cannot be measured directly is the excitation power (coherent scattering of the laser at high powers prevents referencing to a saturation value). Thus, the power has been manually fitted using the measured values of $g^{(2)}(0)$, $g^{(3)}(0)$ and the relative count rates for different excitation conditions.

\section*{Linewidths of detuned systems}

\begin{figure}
  \includegraphics[width=\columnwidth]{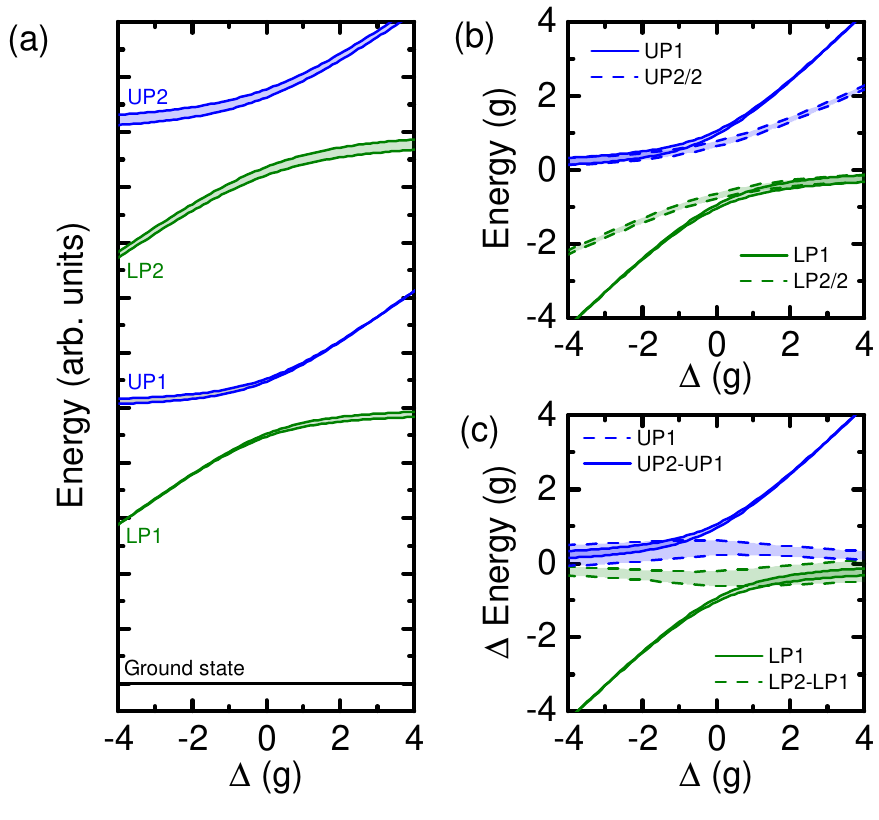}
  \caption{Visualisation of the detuning dependent linewidith: coloured regions indicate the full-width half-max of the transition widths. (a) Schematic illustration of the JC-ladder. (b) Transition energies for an $n$-photon excitation of the $n$-th rung for the first two rungs. (c) Transition energies for subsequent climbs up the ladder. For good visibility of the detuning dependence of the linewidths, in all three subplots the cavity decay rate is depicted as $\kappa = 0.1g$ in order to minimize overlap.}
  \label{figure:S2}
\end{figure}

Using the Liouville equation, dissipation can be introduced into the JC Hamiltonian, resulting in complex energies of the JC system that read \cite{2012_Laussy_Climbing}:
\begin{multline}
E^n_{\pm} = n\omega_a - \frac{\Delta}{2}-i\frac{(2n-1)2\kappa + \gamma}{4} \\ \pm \sqrt{(\sqrt{n}g)^2+\left(\frac{\Delta}{2}-i\frac{2\kappa-\gamma}{4}\right)^2}
\label{eq:En}
\end{multline}
where $E^n_{\pm}$ corresponds to the $n$th rung of the system. The real part of $E^n_{\pm}$ yields the energies of the states whereas the imaginary part yields their linewidths. Interestingly, all four parameters $\kappa$, $\gamma$, $\Delta$ and $g$ contribute to the polariton splittings and linewidths. Additionally, it can be seen from equation \ref{eq:En} that the detuning between different rungs scales with $\sqrt{n}$ while the linewidth scales with $n$. This can be easily understood: The decay rate of a certain rung is proportional to the number of photons involved. Most strikingly, for the case of QDs and photonic crystal cavities with $\kappa \gg \gamma$, the linewidth of a given polariton branch closer to the bare QD (bare cavity) at nonzero detuning is smaller (larger) than at resonance. This is particularly important for photon-blockade phenomena that rely on the absence of an energetic overlap between transitions from the crystal ground state to the first rung and transitions from the first rung to higher rungs. Therefore, detuning the cavity and QD not only increases the energy difference between transitions involving different rungs but also decreases the linewidth of the polariton branches closer to the bare QD transition, thus further enhancing the fidelity of photon-blockade. The dependence of the transition linewidths on the detuning is visualised in figure \ref{figure:S2}. The figure presents in panel (a) the JC ladder, in (b) the transition energies for an $n$-photon excitation of the $n$-th rung and in (c) the transition energies for subsequent excitations. Here, the linewidths are indicated by the coloured areas. Note that due to the large overlap of some transitions for this figure, we took the artistic liberty to colour with the artificially low cavity decay rate of $\kappa = 0.1 \, g$, corresponding to a strongly coupled system with state-of-the-art parameters. While in our experiments the cavity decay rate is larger, the same dependence on the QD-cavity detuning applies.

\section*{Emission from detuned systems}
It is important to note that also for detuned strongly coupled systems the emission from the emitter-like polariton branch occurs almost exclusively through the cavity mode. This can be understood in the following way: As the emitter is embedded in a photonic band-gap material, its emission into free-space modes through non-cavity channels is strongly Purcell suppressed. For the photonic crystal geometries used in our work, measured values of this emission rate are suppressed to approximately one tenth of their bulk value \cite{Englund2005, Kaniber2008}. As a result of this extreme time-scale disparity, the emission of the emitter through the detuned cavity mode obtained from equation \ref{eq:En} dominates for detunings up to $\sim 60\, g$. This value is much larger than the detunings of a few $g$ as investigated in our letter. Therefore, the emission occurs almost exclusively through the cavity. \\

\begin{figure}
  \includegraphics[width=\columnwidth]{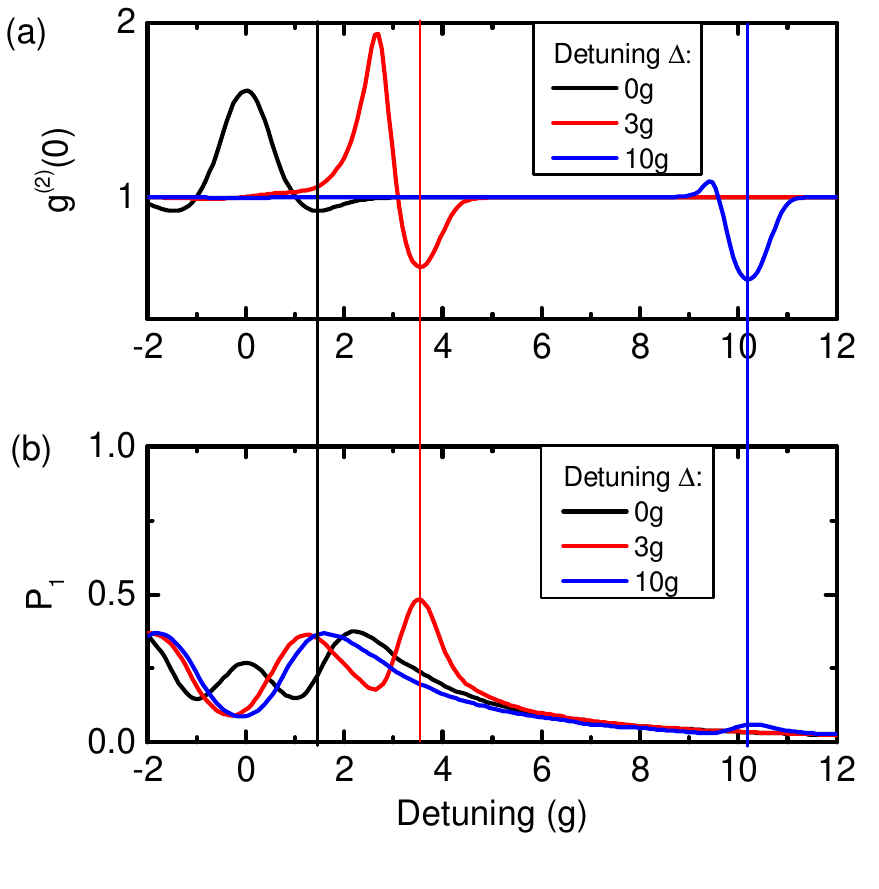}
  \caption{Simulated (a) $g^{(2)}(0)$ and (b) $P_1$ as a function of the laser detuning for a QD-cavity detuning of $0 \, g$ (black), $3 \, g$ (red) and $10 \, g$ (blue).	The positions of the minimum $g^{(2)}(0)$ are indicated by vertical lines.
	}
  \label{figure:S3}
\end{figure}

\section*{Optimum detuning for efficient single-photon generation}
In figure 3 of the main text we have compared the probability of finding $n=0-3$ photons in a pulse transmitted through the cavity. In particular, we found that the probability for obtaining a single photon increases when increasing the detuning from $0 \, g$ to $3 \, g$. However, as the oscillator strength of the QD-like polariton branch decreases with increasing detuning as discussed above, for too large detunings the probability of finding a single photon in the output decreases. To visualise this, we present in figure \ref{figure:S3}a $g^{(2)}(0)$ and in \ref{figure:S3}b $P_1$ as a function of the laser detuning for a QD-cavity detuning of $0 \, g$ (black), $3 \, g$ (red) and $10 \, g$ (blue). While for $\Delta = 10 \, g$ $g^{(2)}(0)$ is close to the value obtained at $\Delta = 3 \, g$, $P_1$ (and thus the efficiency for single-photon generation) decreases to one eight of the value at $\Delta = 3 \, g$. This clearly demonstrates the existence of an \emph{optimum detuning} for efficient single-photon generation.

\begin{figure}
  \includegraphics[width=\columnwidth]{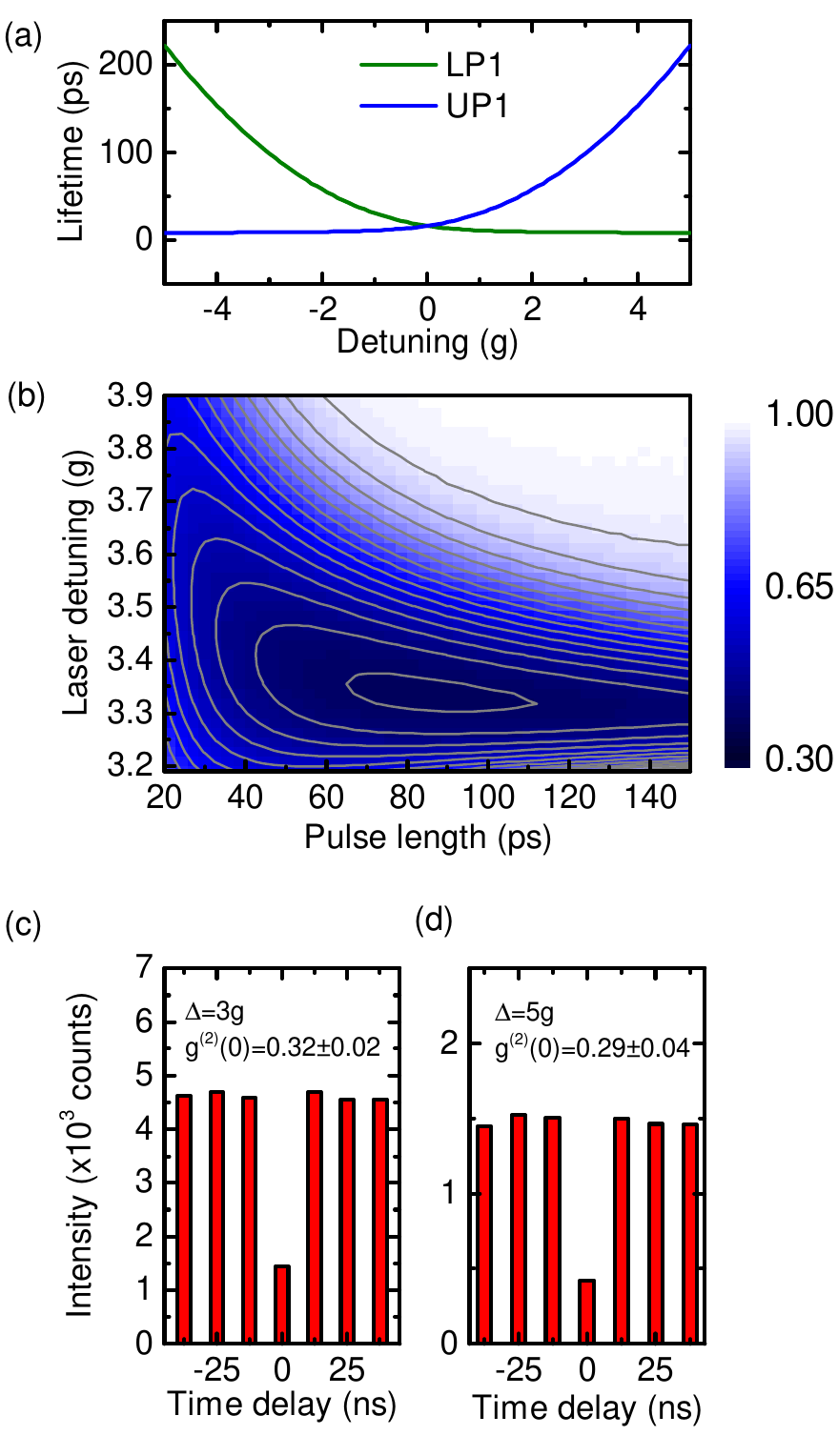}
  \caption{(a) Radiative polariton lifetime as a function of the QD-cavity detuning, as calculated using the measured values of $\kappa$ and $g$. (b) Simulation of $g^{(2)}(0)$ as a function of the pulse length and laser detuning for a cavity QD detuning of $\Delta=3g$. (c-d) Examples of measurements of $g^{(2)}(0)$ using $70\,ps$ long pulses and a detuning of (c) $\Delta = 3g$ and (d) $\Delta = 5g$.
	}
  \label{figure:S4}
\end{figure}

\section*{Influence of the pulse length}
Due to a short cavity lifetime that is much shorter than the timing jitter of the single photon counters, measurements of $g^{(2)}(0)$ can only be performed with pulsed excitation. As discussed in the main text, the choice of pulse length forces a compromise between frequency resolution (reducing the overlap of different rungs) and re-excitation of the system. In other words, if the laser pulses are too long the system will be re-excited during the interaction with a single pulse, reducing the non-classical character of the transmitted light. On the other hand, pulses with a shorter duration are spectrally broader, resulting in a larger overlap with higher rungs. The dependence of the radiative polariton lifetimes on the QD-cavity detuning is calculated by using equation \ref{eq:En} with the measured values of $\kappa$ and $g$ (figure \ref{figure:S4}a). With increasing detuning between QD and cavity, the lifetime of the emitter-like (cavity-like) polariton branch increases (decreases). Therefore, in order to obtain the strongest photon-blockade the pulse length has to be chosen according to the detuning. To demonstrate the existence of an optimum pulse length for a given detuning between QD and cavity, we present in figure \ref{figure:S4}b simulations of $g^{(2)}(0)$ as a function of the pulse length and laser detuning for a QD-cavity detuning of $\Delta = 3g$. With longer excitation pulses the photon-blockade dip narrows and $g^{(2)}(0)$ decreases, reaching its lowest values for pulse lengths of $70-110 \, ps$. As mentioned in the main text, for detunings of $\Delta = 3-5 \, g$ we were able to obtain lower values of $g^{(2)}(0)$ using a pulse duration of $70 \, ps$ instead of $30 \, ps$. Examples of these measurements are presented in figure \ref{figure:S4}c and \ref{figure:S4}d.

\end{document}